\documentclass[twocolumn,amsmath,amssymb]{revtex4}
\usepackage[dvips]{graphicx}
\usepackage{dcolumn}
\usepackage{bm}

\begin{document}

\title[]
{Some remarks on exact wormhole solutions}
\author{Peter K.\,\,F. Kuhfittig}
\address{Department of Mathematics\\
Milwaukee School of Engineering\\
Milwaukee, Wisconsin 53202-3109 USA}
%\date{\today}

\begin{abstract}
\noindent
Exact wormhole solutions, while eagerly saught after, often have
the appearance of being overly specialized or highly artificial.
A case for the possible existence of traversable wormholes would
be more compelling if an abundance of solutions could be found.
It is shown in this note that for many of the wormhole
geometries in the literature, the exact solutions obtained
imply the existence of large sets of additional solutions.

\vspace{12pt}
\noindent
PAC numbers: 04.20.Jb, 04.20.Gz
\newline Key words: wormholes, exact solutions
\end{abstract}

\maketitle

\section{Introduction}\label{S:Introduction}
\noindent
One of the fundamental problems in the general theory of relativity is
finding exact solutions of the Einstein field equations \cite{jB06}.
An example is the Schwarzschild solution, which proved to be a major
milestone.  Similarly, finding exact solutions of wormhole geometries
has occupied many researchers.  The difference is that in this setting
exact solutions are hardly a virtue if they have the appearance of
being artificial or, at best, overly specialized.  It is shown in
this note that exact solutions may be used to show the existence
of entire families of solutions, which in turn helps strengthen the
case for the possible existence of traversable wormholes.
%END OF SECTION

\section{Wormhole solutions}
\noindent
A wormhole may be defined as a handle or tunnel in the spacetime
topology linking different universes or widely separated regions of our
own Universe \cite{MT88}.  A general line element is the following:
\begin{equation}\label{E:line1}
  ds^2=-e^{2\Phi(r)}dt^2+e^{2\Lambda(r)}dr^2
      +r^2(d\theta^2+\text{sin}^2\theta\,d\phi^2),
\end{equation}
where $\Phi(r)\rightarrow 0$ and $\Lambda(r) \rightarrow 0$ as
$r \rightarrow \infty$, usually referred to as \emph{asymptotic
flatness}.  We also assume that $\Phi$ and $\Lambda$
have continuous derivatives in their respective domains.  The
function $\Phi=\Phi(r)$ is called the \emph{redshift function},
which must be everywhere finite to prevent an event horizon.
The function $\Lambda=\Lambda(r)$ is  related to the
\emph{shape function} $b(r)=r(1-e^{-2\Lambda(r)})$, that is,
$e^{2\Lambda(r)}=1/[1-b(r)/r]$. The minimum radius $r=r_0$ is the
\emph{throat} of the wormhole, where $b(r_0)=r_0$.  Also needed is
$b'(r_0)<1$, usually referred to as the \emph{flare-out condition}.
As a consequence, $\Lambda(r)$ has a vertical asymptote at
$r=r_0$: $\lim_{r \to r_0+}\Lambda(r)=+\infty$.  For the redshift
function we also have the additional requirement that
$\Phi'(r_0)$ be finite.  Finally, to hold a wormhole open, the null
energy condition must be violated \cite{MT88}.

The components of the Einstein tensor in the orthonormal frame are
listed next \cite{pK08}.
\begin{equation}\label{E:Einstein1}
  G_{\hat{t}\hat{t}}=\frac{2}{r}e^{-2\Lambda(r)}\Lambda'(r)
    +\frac{1}{r^2}\left(1-e^{-2\Lambda(r)}\right),
\end{equation}
\begin{equation}\label{E:Einstein2}
   G_{\hat{r}\hat{r}}=\frac{2}{r}e^{-2\Lambda(r)}\Phi'(r)
   -\frac{1}{r^2}\left(1-e^{-2\Lambda(r)}\right),
\end{equation}
\begin{multline}\label{E:Einstein3}
   G_{\hat{\theta}\hat{\theta}}=G_{\hat{\phi}\hat{\phi}}\\
   =e^{-2\Lambda(r)}\left(\Phi''(r)-\Phi'(r)\Lambda'(r)+
   [\Phi'(r)]^2\phantom{\frac{1}{r}}\right.\\
    \left. +\frac{1}{r}\Phi'(r)-\frac{1}{r}\Lambda'(r)\right).
\end{multline}

The case that will be examined in some detail is a wormhole supported
by phantom energy, whose equation of state (EoS) is $p=-K\rho$, $K>1$.
For this case, $\rho+p=\rho(1-K)<0$, in violation of the null energy
condition.

From the Einstein field equations $G_{\hat{\alpha}\hat{\beta}}=
8\pi T_{\hat{\alpha}\hat{\beta}}$ and Eqs. (\ref{E:Einstein1})
and (\ref{E:Einstein2}), we now get
\begin{multline*}
\frac{1}{8\pi}\left[\frac{2}{r}e^{-2\Lambda(r)}\Phi'(r)-\frac{1}{r^2}
   \left(1-e^{-2\Lambda(r)}\right)\right]\\
    =-K\frac{1}{8\pi}\left[\frac{2}{r}e^{-2\Lambda(r)}\Lambda'(r)
   +\frac{1}{r^2}\left(1-e^{-2\Lambda(r))}\right)\right].
\end{multline*}
Solving for $\Lambda'(r)$, we have
\begin{equation}\label{E:Lambda1}
   \Lambda'(r)=\frac{1}{K}\left[-\Phi'(r)-\frac{1}{2r}
     \left(e^{2\Lambda(r)}-1\right)(K-1)\right].
\end{equation}
Observe that $\lim_{r\to r_0+}\Lambda'(r)=-\infty$.  Otherwise
$\Lambda'(r)$ is continuous in the interval $(r_0, \infty)$.

The continuity assumptions play a critical role: thus for $a>r_0$,
the integral $\int^b_{a}e^{\Lambda(r)}dr$ exists since
$\Lambda(r)$ is continuous.  Now from the line element
(\ref{E:line1}), the proper radial distance $\ell(r)$ from the
throat $r=r_0$ to any point away from the throat is given by
\begin{equation}
  \ell(r)=\int^r_{r_0}e^{\Lambda(r')}dr'.
\end{equation}
For an ``arbitrary" $\Lambda(r)$, this integral is not likely to be
finite, given the asymptotic behavior of $\Lambda(r)$.  For certain
choices of $\Phi(r)$, however, $\ell(r)$ does exist, thereby
producing a traversable wormhole.  Examples are the exact solutions
for $\Lambda(r)$ obtained by choosing the following redshift
functions: $\Phi(r)\equiv \text{constant}$ \cite{fL05} and
$\Phi(r)=\text{ln}\frac{k}{r}$ \cite{oZ05, pK06}.

Next, let us suppose that $\Lambda_1(r)$ is one of the above exact
solutions [corresponding to some $\Phi(r)$], so that
\[
   \ell_1(r)=\int^r_{r_0}e^{\Lambda_1(r')}dr'
\]
is finite.  Given that $\Phi'(r)$ is finite by assumption, we deduce
from Eq.~(\ref{E:Lambda1}) that
\begin{equation}\label{E:Lambda2}
  q_1(r)=
  \Lambda'_1(r)+\frac{1}{K}\frac{1}{2r}
   \left(e^{2\Lambda_1(r)}-1\right)(K-1)
\end{equation}
is also finite.  To generate a new solution, we let $\epsilon=
\epsilon(r)$ be a function with a continuous derivative
satisfying the conditions $\epsilon(r_0)=\epsilon'(r_0)=0$
and $\epsilon(r)\rightarrow 0$ as $r\rightarrow \infty$.
Define $\Lambda_2(r)=\Lambda_1(r)+\epsilon(r)$.  Then by
Eq.~(\ref{E:Lambda2})
\begin{equation}\label{E:Lambda3}
  q_2(r)=
  \Lambda'_2(r)+\frac{1}{K}\frac{1}{2r}
   \left(e^{2\Lambda_2(r)}-1\right)(K-1)
\end{equation}
is finite.  Now consider the equation
\begin{multline}\label{E:Lambda1modified}
   \Lambda_2'(r)=\\\frac{1}{K}\left[-[\Phi'(r)+\eta'(r)]
  -\frac{1}{2r}\left(e^{2\Lambda_2(r)}-1\right)
   (K-1)\right]
\end{multline}
for some finite differentiable function $\eta=\eta(r)$.
Writing Eq. (\ref{E:Lambda1modified}) in the form
\begin{multline}\label{E:Lambda4a}
   \Lambda_1'(r)+\epsilon'(r)=\\
   \frac{1}{K}\left[-[\Phi'(r)+\eta'(r)]
  -\frac{1}{2r}\left(e^{2\Lambda_1(r)+2\epsilon(r)}
  -1\right) (K-1)\right],
\end{multline}
we see that, by comparison to Eq. (\ref{E:Lambda1}),
$\eta'(r)$ must satisfy the conditions $\eta'(r_0)=0$
and $\eta(r)\rightarrow 0$ as $r\rightarrow \infty$,
thereby retaining the asymptotic flatness.  The
function $\eta'(r)$ is defined only implicitly in
Eq.~(\ref{E:Lambda4a}), but since we are strictly
interested in the \emph{existence} of new solutions,
it is sufficient to note that the formal relationship
$\eta'(r)=-\Phi'(r)-Kq_2(r)$ implies that $\eta'(r)$
exists and is sectionally continuous, so that
$\eta=\eta(r)$ exists, as well.

We conclude that $\Lambda_2(r)$ is a solution to
Eq. (\ref{E:Lambda1}) corresponding to some finite
redshift function $\Phi(r)+\eta(r)$.  Moreover,
\begin{equation}
  \ell_2(r)=\int^r_{r_0}e^{\Lambda_2(r')}dr'
\end{equation}
is finite since $\epsilon(r_0)=0$.  The result is the existence
of an entire family of solutions, one for each
$\epsilon=\epsilon(r)$.

Similar arguments can be used to obtain new classes of solutions
from exact solutions of wormholes supported by Chaplygin and
generalized Chaplygin gas \cite{fL06}, modified Chaplygin
gas \cite{CB07}, and even van der Waals quintessence \cite {fL07}.
For the case of a generalized Chaplygin gas, the EoS
is $p=-A/\rho^{\alpha}$, $0<\alpha\le 1$, and where $A$ is a
constant \cite{fL06}.  After substituting and solving for
$\Lambda'(r)$, we get \cite{pK08}
\begin{multline}\label{E:Lambda4}
  \Lambda'(r)=\\ \frac{\frac{1}{2}(8\pi)^{1+1/\alpha}r(r^2)^{1/\alpha}
   A^{1/\alpha}e^{2\Lambda(r)}}{\left[1-e^{-2\Lambda(r)}
     -2re^{-2\Lambda(r)}\Phi'(r)\right]^{1/\alpha}}
       -\frac{1}{2r}\left(e^{2\Lambda(r)}-1\right).
\end{multline}
According to Ref. \cite{fL06}, there is an exact solution for
$\Phi'\equiv 0$, making $\ell(r)$ finite.  In analogous manner,
we now let  $\Lambda_1(r)$ be an exact solution of
Eq.~(\ref{E:Lambda4}).  Then $\Lambda_2(r)+\epsilon(r)$ is a
new solution, corresponding to some $\Phi(r)+
\eta(r)$.  (Assume that $\epsilon(r)$ and $\eta(r)$
satisfy the same conditions as before.)  The resulting
proper distance $\ell_2(r)$ is again finite.

An example that illustrates the problem of generalizing an exact
solution even better comes from the modified Chaplygin gas model, whose
EoS is $p=A\rho-B/\rho^{\alpha}$, $0<\alpha\le 1$, and where $A$ and $B$ are
constants.  This time we begin with $\Lambda(r)$ and determine $\Phi(r)$.

According to Ref. \cite{CB07}, one may choose $b(r)=r_0+d(r-r_0)$, where $d$
must be less than unity to meet the flare-out condition.  The shape function
$b(r)$ determines $\Lambda(r)$, as well as $\Phi(r)$ \cite{CB07}:
\begin{multline}
   \Phi(r)= \phi_0+\frac{1+Ad}{2(1-d)}\text{ln}|r-r_0|
    -\frac{1}{2}\text{ln}\,r\\-\frac{32\pi^2B}{d(1-d)}
         r^4_0\,\text{ln}|r-r_0|\\-\frac{32\pi^2B}{d(1-d)}
     \left[\frac{1}{4}(r-r_0)^4+4r_0^3(r-r_0)\right.\\
     \left.+3r_0^2(r-r_0)^2+\frac{4}{3}r_0(r-r_0)^3\right].
\end{multline}
This $\Phi(r)$ is not finite unless $A$ has the (corrected) value
\begin{equation}\label{E:A}
   A=\frac{1}{d^2}(64\pi^2Br_0^4-d).
\end{equation}
\emph{Remark:} Since $A$ is part of the EoS, it would be more realistic
to state the condition as follows: determine $d$\, $(0<d<1)$, if it
exists, so that Eq.~(\ref{E:A}) is satisfied.

For such a choice of $d$, we find that
\begin{multline}
   \Phi(r)= \phi_0
    -\frac{1}{2}\text{ln}\,r-\frac{32\pi^2B}{d(1-d)}
     \left[\frac{1}{4}(r-r_0)^4\right.\\
       \left.+4r_0^3(r-r_0)
     +3r_0^2(r-r_0)^2+\frac{4}{3}r_0(r-r_0)^3\right],
\end{multline}
which has a  well defined and continuous derivative.  If we now
substitute Eqs. (\ref{E:Einstein1}) and (\ref{E:Einstein2})
in the EoS and solve for $\Phi'(r)$, we get
\begin{multline}
   \Phi'(r)=\frac{1}{2r}\left(e^{2\Lambda(r)}-1\right)\\
   +A\left[\Lambda'(r)+\frac{1}{2r}\left(e^{2\Lambda(r)}-1
        \right)\right]\\
   -\frac{(8\pi)^{1+\alpha}B(\frac{r}{2})e^{2\Lambda(r)}}
  {\left[\frac{2}{r}e^{-2\Lambda(r)}\Lambda'(r)
     +\frac{1}{r^2}\left(1-e^{-2\Lambda(r)}\right)\right]^{\alpha}}.
\end{multline}
Since the left side is well behaved (due to the choice of $d$),
the right side is also well behaved in the sense of being finite and
sectionally continuous.  As before, we can replace $\Lambda(r)$ by
$\Lambda(r)+\epsilon(r)$, where, once again, $\epsilon=\epsilon(r)$
has a continuous derivative with $\epsilon(r_0)=\epsilon'(r_0)=0$
and $\epsilon(r)\rightarrow 0$ as $r\rightarrow\infty$.  Now denote
the left side by $\Phi'_1(r)$.  Since $\Phi'_1(r)$ is also sectionally
continuous, $\Phi_1(r)$ will exist, again showing the existence of
an entire family of solutions.
%END OF SECTION
\phantom{a}

\phantom{a}

\phantom{a}

\phantom{a}

\phantom{a}
\section{Conclusion}
\noindent
We have shown in this note that exact solutions of certain wormhole
geometries, such as traversable wormholes supported by phantom energy,
Chaplygin and generalized Chaplygin gas, or modified Chaplygin gas,
imply the existence of entire families of additional solutions.
Having an abundance of solutions is highly desirable since exact
solutions often have the appearance of being overly specialized
or even artificial.

From a physical standpoint, the proliferation of solutions, given
the various equations of state, would greatly increase the probability
of wormholes' having formed naturally.  In particular, it is
argued in Ref. \cite{pK08a} that the EoS had very likely crossed the
``phantom divide" some time in the past, so that wormholes could have 
formed naturally.  In approaching the present dark-energy phase, 
such wormholes would have formed
event horizons, thereby becoming black holes or (possibly)
quasi-black holes.  This outcome provides at least a \emph{possible}
explanation for the abundance of black holes and the complete lack
of wormholes.

\end{document}